\documentclass[a4paper,11pt]{article}

\usepackage{latexsym}
\usepackage{amssymb}
\usepackage{amsfonts}
\usepackage{amsmath}
\usepackage{amsthm}
\usepackage{indentfirst}
\usepackage{epsfig}
\usepackage{xypic}
\usepackage[english]{babel}

\usepackage{fullpage}
\usepackage{graphicx}
\usepackage{pstricks}
\usepackage{caption}
\usepackage{pst-plot, pst-node, pst-text, pst-tree}
\usepackage{rotating}
\usepackage{epic, eepic}
\usepackage{color}
\usepackage{indentfirst}
\usepackage{mathrsfs}
\usepackage{manfnt}

\def\sdwys #1{\xHyphenate#1$\wholeString}
\def\xHyphenate#1#2\wholeString {\if#1$%
\else\say{\ensuremath{#1}}\hspace{2pt}%
\takeTheRest#2\ofTheString
\fi}
\def\takeTheRest#1\ofTheString\fi
{\fi \xHyphenate#1\wholeString}
\def\say#1{\begin{turn}{-90}\ensuremath{#1}\end{turn}}

\newenvironment{threestacks}
{
	\begin{footnotesize}
	\psset{xunit=0.0355in, yunit=0.0355in, linewidth=0.02in}
	\begin{pspicture}(0,0)(35,20)
	\psline{c-c}(-5,15)(10,15)(10,0)(15,0)(15,15)(20,15)(20,0)(25,0)(25,15)(30,15)(30,0)(35,0)(35,15)(50,15)
	\rput[l](-5,12.5){\mbox{output}}
	\rput[r](50,12.5){\mbox{input}}
	\rput[h](32.5,-2.5){$D_1$}
	\rput[h](22.5,-2.5){$D_2$}
	\rput[h](12.5,-2.5){\mbox{I}}
}
{
	\end{pspicture}
	\end{footnotesize}
}

\newcommand{\fillthreestacks}[5]{%
	\rput[l](-5,17.5){\ensuremath{#1}}
	\rput[c](12.6, 7.5){\begin{sideways}{\sdwys{#2}}\end{sideways}}
	\rput[c](22.6, 7.5){\begin{sideways}{\sdwys{#3}}\end{sideways}}
    \rput[c](32.6, 7.5){\begin{sideways}{\sdwys{#4}}\end{sideways}}
	\rput[r](50,17.5){\ensuremath{#5}}
}

\newenvironment{threestacksBig}
{
	\begin{footnotesize}
	\psset{xunit=0.0355in, yunit=0.0355in, linewidth=0.02in}
	\begin{pspicture}(0,0)(35,20)
	\psline{c-c}(-5,15)(5,15)(5,0)(20,0)(20,15)(30,15)(30,0)(45,0)(45,15)(55,15)(55,0)(70,0)(70,15)(85,15)
	\rput[l](-7,12.5){\mbox{output}}
	\rput[r](81,12.5){\mbox{input}}
	\rput[h](62.5,-2.5){$D_1$}
	\rput[h](37.5,-2.5){$D_2$}
	\rput[h](12.5,-2.5){\mbox{I}}
}
{
	\end{pspicture}
	\end{footnotesize}
}

\newcommand{\fillthreestacksBig}[5]{%
	\rput[l](-5,17.5){\ensuremath{#1}}
	\rput[c](12.6, 7.5){\begin{sideways}{\sdwys{#2}}\end{sideways}}
	\rput[c](37.6, 7.5){\begin{sideways}{\sdwys{#3}}\end{sideways}}
    \rput[c](62.6, 7.5){\begin{sideways}{\sdwys{#4}}\end{sideways}}
	\rput[r](85,17.5){\ensuremath{#5}}
}

\newenvironment{threestacksVerybig}
{
	\begin{footnotesize}
	\psset{xunit=0.0355in, yunit=0.0355in, linewidth=0.02in}
	\begin{pspicture}(-30,0)(0,20)
	\psline{c-c}(-5,15)(5,15)(5,-10)(20,-10)(20,15)(30,15)(30,-10)(45,-10)(45,15)(55,15)(55,-10)(70,-10)(70,15)(110,15)
	\rput[l](-7,12.5){\mbox{output}}
	\rput[r](90,12.5){\mbox{input}}
	\rput[h](62.5,-12.5){$D_1$}
	\rput[h](37.5,-12.5){$D_2$}
	\rput[h](12.5,-12.5){\mbox{I}}
}
{
	\end{pspicture}
	\end{footnotesize}
}

\newcommand{\fillthreestacksVerybig}[5]{%
	\rput[l](-5,17.5){\ensuremath{#1}}
	\rput[c](12.6, 2.5){\begin{sideways}{\sdwys{#2}}\end{sideways}}
	\rput[c](37.6, 2.5){\begin{sideways}{\sdwys{#3}}\end{sideways}}
    \rput[c](62.6, 2.5){\begin{sideways}{\sdwys{#4}}\end{sideways}}
	\rput[r](110,17.5){\ensuremath{#5}}
}

\newtheorem{theorem}{Theorem}[section]
\newtheorem{prop}[theorem]{Proposition}
\newtheorem{lemma}[theorem]{Lemma}
\newtheorem{cor}[theorem]{Corollary}

\newtheoremstyle{definizione} 
    {\topsep}                    
    {\topsep}                    
    {}                           
    {}                           
    {\bfseries}                   
    {.}                          
    {.5em}                       
    {}  

\theoremstyle{definizione}
\newtheorem{remark}[theorem]{Remark}

\newcommand{\cvd}{\hfill $\blacksquare$\bigskip}

\date{}

\author{Giulio Cerbai\thanks{Dipartimento di Matematica e Informatica ``U.
Dini", University of Firenze, Firenze, Italy,
\tt{giulio.cerbai@unifi.it, lapo.cioni@unifi.it, luca.ferrari@unifi.it}}
\and Lapo Cioni$^{\dag}$ \and Luca Ferrari$^{\dag}$}

\title{Stack Sorting with Increasing and Decreasing Stacks\footnote{G. C. and L. F. are members of the INdAM Research group GNCS; they are partially supported by INdAM - GNCS 2019 project ``Studio di propriet\'a combinatoriche di linguaggi formali ispirate dalla biologia e da strutture bidimensionali" and by a grant of the "Fondazione della Cassa di Risparmio di Firenze" for the project "Rilevamento di pattern: applicazioni a memorizzazione basata sul DNA, evoluzione del genoma, scelta sociale".}}

\begin{document}

\maketitle

\begin{abstract}
We introduce a sorting machine consisting of $k+1$ stacks in series:
the first $k$ stacks can only contain elements in decreasing order from top to bottom, while the last one has the opposite restriction.
This device generalizes \cite{SM}, which studies the case $k=1$.
Here we show that, for $k=2$,
the set of sortable permutations is a class with infinite basis, by explicitly finding an antichain of minimal nonsortable permutations.
This construction can easily be adapted to each $k \ge 3$.
Next we describe an optimal sorting algorithm, again for the case $k=2$.
We then analyze two types of left-greedy sorting procedures, obtaining complete results in one case and only some partial results in the other one.
We close the paper by discussing a few open questions.
\end{abstract}

\bigskip

\section{Introduction}

The problem of sorting a permutation using a stack was first introduced by Knuth \cite{KN} in the 1960s;
in its classical formulation, the aim is to sort a permutation using a first-in/last-out device.
As it is well known,
in this case a permutation $\pi=\pi_1 \cdots \pi_n$ is sortable if and only if there do not exist three indices $i<j<k$ such that $\pi_k < \pi_i < \pi_j$. In the language of permutation patterns, we say that the set of sortable permutations is a \textit{class} with basis $\left\lbrace 231 \right\rbrace$,
meaning that each of these permutations cannot contain the pattern $231$ as a subpermutation;
a class is a downset in the permutation pattern poset and each class is determined by the minimal elements in its complement,
which form its \textit{basis}.
Recall that the set of permutations can be partially ordered by means of the relation of ``being a pattern",
and we write $\sigma \leq \pi$ to mean that $\sigma$ is a pattern of $\pi$.
The resulting poset is called the \emph{permutation pattern poset},
and a downset (i.e., a subset closed by going downwards) of the permutation pattern poset is usually called a class.
For the basics on permutation patterns in combinatorics and computer science, we refer to \cite{Bona}.

More generally (see \cite{TA}), one can consider a network of sorting devices, each of which is represented as a node in a directed graph;
when there is an arc from node $S$ to node $T$, the machine is allowed to pop an element from $S$ and push it into $T$;
if we mark two distinct vertices as the input and the output,
then the sorting problem consists of looking for a sequence of operations that allows us to move a permutation from the input to the output,
finally obtaining the identity permutation.

In this framework, some of the typical problems are the following:

\begin{itemize}
\item characterize the permutations that can be sorted by a given network;
\item enumerate sortable permutations with respect to their length;
\item if the network is too complex, find a specific algorithm that sorts ``many" input permutations and characterize such permutations.
\end{itemize}

Concerning the last problem, note that, for a given network of devices, although the set of sortable permutations forms a class in general,
this is not true anymore if one chooses a specific sorting strategy;
this approach leads in general to more complicated characterizations which involve other kinds of patterns
(as it happens, for instance, for West 2-stack-sortable permutations \cite{WE}).

Although it is very hard to obtain interesting results for large networks,
a lot of work has been done for some particular, small networks (see \cite{BO} for a dated survey, or \cite{K} for a more recent one);
in this work we restrict our attention to the case of stacks connected in series,
with the restriction that the elements are maintained inside each stack either in increasing or in decreasing order.
Our starting point is \cite{SM},
where Rebecca Smith proved that the permutations sorted by a decreasing stack followed by an increasing one form a class with basis
$\left\lbrace 3241,3142 \right\rbrace$.
In the present paper, we try to find some information on what happens when we add more decreasing stacks in front.
Our first result is that
the device having two decreasing stacks followed followed by an increasing one does not have a finite basis.
Our proof can be easily adapted to show the same property for any number of decreasing stacks in front.
Next, we provide an optimal algorithm to sort permutations, again in the case of two decreasing stacks followed by an increasing one.
Our algorithm is optimal in the sense that it is able to sort all sortable permutations.
Finally, we select a couple of (greedy) strategies and we prove that one of them can be studied in a very neat way,
whereas the other one seems to be too difficult to allow a simple description of sortable permutations
in terms of patterns, even including generalized versions of them.

\section{Many decreasing stacks followed by an increasing one.}

Generalizing the approach of \cite{SM},
here we will consider a sorting device made by $k$ decreasing stacks in series, denoted by $D_1 ,\dots ,D_k$, followed by an increasing stack $I$.
Recall that ``decreasing" (resp., ``increasing") stack means
that the elements inside the stack have to be in decreasing (resp., increasing) order from top to bottom.
When $k=0$, we just have a single increasing stack, so we obtain the usual Stacksort procedure.
When $k=1$, we obtain exactly the $\mathfrak{DI}$ machine described in \cite{SM}. In the sequel we denote our machine with $\mathfrak{D}^k \mathfrak{I}$.

The $\mathfrak{D}^k \mathfrak{I}$ machine can perform the following operations:
\begin{itemize}
\item $d_0$: push the next element of the input permutation into the first decreasing stack $D_1$;
\item $d_i$, for $i=1,\dots ,k-1$: pop an element from $D_i$ and push it into the next decreasing stack $D_{i+1}$;
\item $d_k$: pop an element from the last decreasing stack $D_k$ and push it into the increasing stack $I$;
\item $d_{k+1}$: pop an element from the increasing stack $I$ and output it (by placing it on the right of the list of elements that have already been output).
\end{itemize}

Notice that each operation can be performed only if it does not violate the restrictions of the stacks;
in this case, we call it a \textit{legal} operation.
For the special case of the operation $d_{k+1}$,
we will assume that $d_{k+1}$ is legal both if we are pushing into the output the smallest among the elements not already in the output
and if all the other operations are not legal.

\begin{remark}\label{Not_sort}
If an occurrence of the pattern $231$ is pushed into the last stack $I$, then the input permutation cannot be sorted.
Moreover, this is the only situation that corresponds to a failure in the sorting procedure.
This is a consequence of the classical result of Knuth \cite{KN}, where in fact the only stack is used exactly as if it were increasing.
\end{remark}

For any given $k$, we are now interested in characterizing the set
$$Sort_k = \{ \pi\in S\, |\, \mbox{ there is a sequence of legal operations of the $\mathfrak{D}^k \mathfrak{I}$ machine that sorts } \pi \}.$$

If $\pi \in Sort_k$, we say that $\pi$ is $k$-sortable.
Notice that we are using the sorting machine in the most general setting,
so using a standard argument it is easy to show that $Sort_k$ is a class for every $k$.
The natural way to describe $Sort_k$ is therefore to understand its basis.
Here we show that, even when $k=2$, the basis of $Sort_k$ is infinite,
by explicitly finding an infinite antichain of permutations which are not $2$-sortable and are minimal with respect to the pattern ordering.
The construction of the infinite antichain described in the next theorem can be easily adapted to every $k \ge 2$.
The software \emph{PermLab} \cite{AL}, developed by Michael Albert, has been an extremely useful tool to find such an antichain.
This result is in sharp contrast with what happens when $k=1$, which is the case considered in \cite{SM},
where it is shown that the basis is finite (of cardinality 2).
We start by stating some useful lemmas, whose proofs are straightforward.

\begin{lemma}
Let $\pi$ be an input permutation for the $\mathfrak{D}^k \mathfrak{I}$ machine;
if $i<j$ and $\pi_i > \pi_j$, then $\pi_i$ is necessarily pushed into $I$ before $\pi_j$.
In other words, the decreasing stacks $D_1,\ldots ,D_k$ cannot repair inversions.
\end{lemma}

\begin{lemma}\label{not_sort2}
Let $\pi$ be an input permutation for the $\mathfrak{D}^k \mathfrak{I}$ machine and let $a<b<c$ be elements of $\pi$.
Focus on the instant when, during the sorting process, $b$ is pushed into the increasing stack.
Then, if any of the following conditions holds, $\pi$ cannot be sorted anymore:
\begin{enumerate}
\item $c$ is in $D_j$ and $a$ is in $D_k$, with $k \le j$;
\item $c$ is in $D_j$, for some $j$, and $a$ is still in the input;
\item $c$ and $a$ are still in the input, with $a$ following $c$.
\end{enumerate}
\end{lemma}
\emph{Proof.} \quad The previous lemma implies that, if any of the above conditions is satisfied, an occurrence of the pattern $231$ is pushed into the increasing stack,
so $\pi$ cannot be sorted anymore due to Remark \ref{Not_sort}.\cvd

Rephrasing the last lemma,
if we try to sort $\pi$ and, when $b$ is pushed into the increasing stack, one of the listed conditions holds,
then there is no hope to complete the procedure to obtain a sorted output.

\begin{figure}
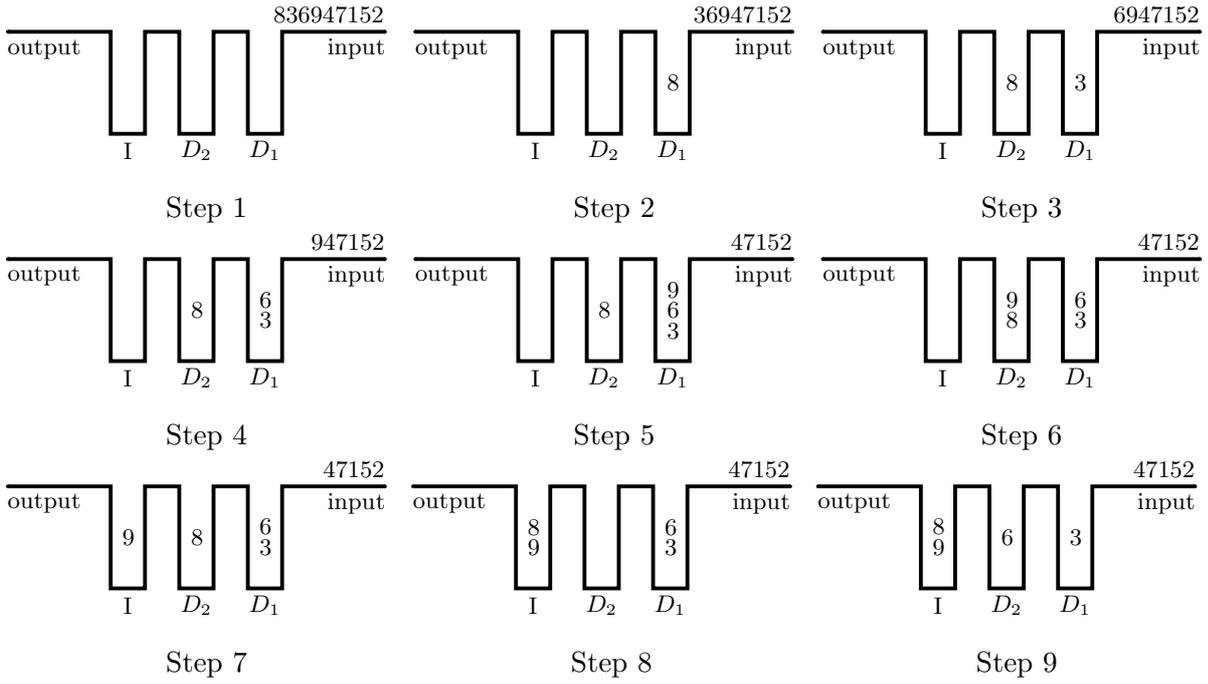

\begin{minipage}[c]{5.2cm}
\begin{threestacks}
\fillthreestacks{}{}{}{}{836947152}
\end{threestacks}

\vspace{0.2cm}

\begin{flushleft}
\hspace{1.5cm} Step 1
\end{flushleft}
\end{minipage}
\begin{minipage}[c]{5.2cm}
\begin{threestacks}
\fillthreestacks{}{}{}{8}{36947152}
\end{threestacks}

\vspace{0.2cm}

\begin{flushleft}
\hspace{1.5cm} Step 2
\end{flushleft}
\end{minipage}
\begin{minipage}[c]{5.2cm}
\begin{threestacks}
\fillthreestacks{}{}{8}{3}{6947152}
\end{threestacks}

\vspace{0.2cm}

\begin{flushleft}
\hspace{1.5cm} Step 3
\end{flushleft}
\end{minipage}
\begin{minipage}[c]{5.2cm}
\begin{threestacks}
\fillthreestacks{}{}{8}{36}{947152}
\end{threestacks}

\vspace{0.2cm}

\begin{flushleft}
\hspace{1.5cm} Step 4
\end{flushleft}
\end{minipage}
\begin{minipage}[c]{5.2cm}
\begin{threestacks}
\fillthreestacks{}{}{8}{369}{47152}
\end{threestacks}

\vspace{0.2cm}

\begin{flushleft}
\hspace{1.5cm} Step 5
\end{flushleft}
\end{minipage}
\begin{minipage}[c]{5.2cm}
\begin{threestacks}
\fillthreestacks{}{}{89}{36}{47152}
\end{threestacks}

\vspace{0.2cm}

\begin{flushleft}
\hspace{1.5cm} Step 6
\end{flushleft}
\end{minipage}
\begin{minipage}[c]{5.2cm}
\begin{threestacks}
\fillthreestacks{}{9}{8}{36}{47152}
\end{threestacks}

\vspace{0.2cm}

\begin{flushleft}
\hspace{1.5cm} Step 7
\end{flushleft}
\end{minipage}
\begin{minipage}[c]{5.2cm}
\begin{threestacks}
\fillthreestacks{}{98}{}{36}{47152}
\end{threestacks}

\vspace{0.2cm}

\begin{flushleft}
\hspace{1.5cm} Step 8
\end{flushleft}
\end{minipage}
\begin{minipage}[c]{5.2cm}
\begin{threestacks}
\fillthreestacks{}{98}{6}{3}{47152}
\end{threestacks}

\vspace{0.2cm}

\begin{flushleft}
\hspace{1.5cm} Step 9
\end{flushleft}
\end{minipage}
\caption{The recursive construction described in Theorem \ref{antichain} with input $\alpha^{(2)}=836947152$ (on the right).
The last step corresponds to input $\alpha^{(1)}=6347152$ after having pushed the first two elements into the machine.}{\label{fig: induction}}
\end{figure}

\begin{theorem}\label{antichain}
For $j \ge0$, define the permutation:
$$\alpha^{(j)}=2j+4,3,\omega^{(j)},1,5,2,$$
where $\omega^{(j)}=2j+2,2j+5,2j,2j+3,2j-2,2j+1,\dots,6,9,4,7$.
Then the set of permutations $\{\alpha^{(j)} \}_{j \ge 0}$ constitutes an infinite antichain in the permutation pattern poset,
each of whose element is not 2-sortable.
Moreover, $\alpha^{(j)}$ is minimal with respect to such a property, i.e. if we remove any element of $\alpha^{(j)}$ we obtain a $2$-sortable permutation.
\end{theorem}
\emph{Proof.}\quad We start by proving (using induction) that $\alpha^{(j)}$ is not 2-sortable, for every $j$.
If $j=0$, it is easy to check that $\alpha^{(0)}=43152$ cannot be sorted using the $\mathfrak{D}^2 \mathfrak{I}$ machine.
Let $j \ge 1$ and $\alpha^{(j)}=\alpha_1 \cdots \alpha_{2j+5}$.
Since $\alpha_1=2j+4 > \alpha_2=3$, $\alpha_1$ has to be pushed into $D_2$ before $\alpha_2$ enters $D_1$.
Notice that the maximum of $\alpha^{(j)}$ is $\alpha_4=2j+5$
and there are elements following it in $\alpha^{(j)}$ which are smaller than both $\alpha_1$ and $\alpha_4$,
so we cannot push $\alpha_1$ into $I$ due to the previous lemma.
Thus the only option we are left with is to push $\alpha_3=2j+2$ into $D_1$ immediately above $\alpha_2$.
Now, the next element of the input is the maximum $\alpha_4$, and of course we can push it through the decreasing stacks and finally into $I$.
Observe that pushing the maximum available element in $I$ is always convenient.
So the second maximum $\alpha_1=2j+4$, which is currently contained in $D_2$, can be pushed into $I$ similarly,
leaving us with just the elements $\alpha_3$ and $\alpha_2$ in $D_1$, with $\alpha_3$ on top.
The next element of the input is $\alpha_5 =2j<\alpha_3$, so pushing $\alpha_3$ into $D_2$ is forced.
Now, getting rid of the two maximal elements of $\alpha^{(j)}$ already pushed into $I$,
notice that we are in the same configuration that arises when processing $\alpha^{(j-1)}$ after considering the first two elements,
so we can conclude that $\alpha^{(j)}$ is not 2-sortable by inductive hypothesis.
An example of the above argument for $j=2$ is shown in \figurename~\ref{fig: induction}.
In passing, we observe that the optimal sorting strategy here would be, at each step,
to push the maximum and second maximum element still available into $I$;
in the general case, this strategy fails since $3$ remains stuck in $D_1$, blocked by a larger element in $D_2$,
until we reach the final portion of $\alpha^{(j)}$.
This crucial remark will be useful in the last part of this proof.

We now prove that $\alpha^{(j)}$ is minimal not 2-sortable.
This can be proved with a case by case analysis, depending on the element we choose to remove.
We show in detail just some of these cases, leaving the remaining ones to the reader.

\begin{itemize}
\item If we remove the first element $\alpha_1=2j+4$, we can push the new first element $\alpha_2=3$ directly into $D_2$;
    from now on, we can follow the sorting procedure outlined above, pushing at each step the maximum and second maximum available elements into $I$.
    However in this case, before processing the three last elements $1,5,2$, we have that  both $3$ and $4$ are in $D_2$,
    whereas in processing $\alpha^{(j)}$ we have $3$ inside $D_1$ and $4$ inside $D_2$.
    Therefore we can now push $1$ into $D_1$ and $5$ into $I$ and finally $4,3,2,1$ in the correct order, as desired.
\item If we remove $\alpha_2=3$, we can sort the resulting permutation using the same procedure,
    this time obtaining a configuration with just $4$ in $D_2$ and $1,5,2$ in the input.
\item Consider the removal of an element $x=\alpha_i$, for some $i=3,\dots,2j+2$.
    In the first part of the sorting procedure, the element $3$ is stuck into $D_1$, similarly to what happens when processing $\alpha^{(j)}$.
    However, as soon as we scan the element that follows $x$ in $\alpha^{(j)}$,
    when we push maximum and second maximum in $I$ we are left for a moment with the stack $D_2$ empty (and just $3$ in $D_1$),
    because we removed the element $x$ that had to occupy $D_2$.
    So we can take advantage of this fact and move $3$ into $D_2$, concluding the sorting procedure as in the previous cases.
\item The removal of the elements $1,5,2$ can be dealt with in a similar way.
\end{itemize}

Thus we have seen that, in any case, removing any element of $\alpha^{(j)}$ results in a 2-sortable permutation,
so $\alpha^{(j)}$ is minimal not 2-sortable.\cvd

\begin{cor}
The basis of $Sort_2$ is infinite, since it contains the infinite antichain $\{\alpha^{(j)} \}_{j \ge 0}$ defined in the previous theorem.
\end{cor}

\begin{remark}
Theorem \ref{antichain} remains true if we permute the elements 1,2,3 of $\alpha^{(j)}$, for every $j$.
\end{remark}

\section{An optimal algorithm for the $\mathfrak{D}^2 \mathfrak{I}$-machine}\label{optimal}

The results of the previous section suggest that it may be very hard to enumerate $k$-sortable permutations when $k\geq 2$.
In the present section, we show that, when $k=2$, we are at least able to design an optimal algorithm, called $\mathcal{D}^2 \mathcal{I}$,
which sorts all 2-sortable permutations.

\bigskip

Algorithm $\mathcal{D}^2 \mathcal{I}$ can be explicitly described as follows:

\begin{enumerate}

\item If $Top(I)$ is the next element to be output, then perform $d_3$.

\item If all the elements contained in $D_1$ and $D_2$ are the next elements to be output, then move them to the output.

\item If each of the previous instructions cannot be executed, perform $d_1$, provided that condition ($\beta$) holds.

\item If each of the previous instructions cannot be executed, perform $d_0$, provided that condition ($\gamma$) holds.

\item If each of the previous instructions cannot be executed, perform $d_2$, provided that condition ($\alpha$) holds.

\item Otherwise, perform $d_3$.

\end{enumerate}

Conditions ($\alpha$), ($\beta$) and ($\gamma$) are the following
(we remark that, when a stack is empty, any statement about it is considered to be true):

\begin{itemize}

\item[($\alpha$)] $Top(D_2 )<Top(I)$.

\item[($\beta$)] $Top(D_2 )<Top(D_1 )$ and $Top(D_1 )<Top(I)$.

\item[($\gamma$)] $Top(D_1 )<Input$, $Input<Top(I)$ and the sequence of elements from $Input$ to the first element larger than $Top(D_2 )$ is increasing.

\end{itemize}

In the sequel, each of the $d_i$'s, for $i=0,1,2,3$, will be called an \emph{operation}, exactly as we did until now.
Instead, each of the six items in the above description of  algorithm $\mathcal{D}^2 \mathcal{I}$ will be called an \emph{instruction}.
Therefore, an instruction of $\mathcal{D}^2 \mathcal{I}$ consists of performing a (legal) operation, provided that some constraints are satisfied.

\bigskip

It is not difficult to realize that instruction 2 of the above algorithm is not essential for its correctness, so in principle we could remove it.
However, in some cases (and in particular in the proof of the optimality) it is convenient to have it.

Algorithm $\mathcal{D}^2 \mathcal{I}$ sets certain priorities between operations, provided that certain conditions are fulfilled.
In general, given any two operations $\tilde{d}$ and $\bar{d}$,
we will use the notation $\tilde{d}\rhd \bar{d}$ to mean that $\tilde{d}$ has higher priority than $\bar{d}$
(and so, if both $\tilde{d}$ and $\bar{d}$ are legal, $\tilde{d}$ is performed).
Moreover, we denote with $^{(\omega)}d$ any operation $d$
which, in order to be performed, has to be legal and also to satisfy an additional constraint $\omega$.

Using these notations,
we can illustrate algorithm $\mathcal{D}^2 \mathcal{I}$ (in which instruction 2 has been removed) with the following chain of priorities:

$$
d_3 \, \rhd \, ^{(\beta)}d_1 \, \rhd \, ^{(\gamma)}d_0 \, \rhd \, ^{(\alpha)}d_2 .
$$

Notice that condition ($\alpha$) is equivalent to saying that operation $d_2$ is legal;
however, for homogeneity's sake, we have preferred to state it explicitly in the description of our algorithm.

\bigskip

\emph{Remarks.}

\begin{enumerate}
\item If, at some point, algorithm $\mathcal{D}^2 \mathcal{I}$ performs instruction 6,
then the input permutation is not sorted at the end of the process,
and this is the only obstruction to the sorting process.
In other words, $\mathcal{D}^2 \mathcal{I}$ sorts a permutations if and only if it never executes instruction 6.
\item To some extent,
algorithm $\mathcal{D}^2 \mathcal{I}$ generalizes Smith's algorithm
for a decreasing stack and an increasing stack in series.
More specifically,
interpreting the first stack of our device as the input container (and so removing the decreasing constraint)
and operation $d_1$ as the input operation,
which insert the current element of the input permutation into the (new) first decreasing stack,
we obtain precisely Smith's algorithm.
\end{enumerate}

The proof of the optimality of our algorithm is not trivial, and requires several steps.
Our first goal is to prove some properties of algorithm $\mathcal{D}^2 \mathcal{I}$.

\begin{lemma}\label{alfa}
At every step, we have $Top (D_2 )<Top (I)$.
\end{lemma}

\emph{Proof.}\quad By induction on the step number.
At the beginning of the sorting process, the statement in the lemma is true since all the stacks are empty.
Now suppose that the statement holds at step $n$, and consider all possible instructions that can be performed:
a simple case-by-case analysis shows that the same inequality is true also at step $n+1$.\cvd

\begin{cor}\label{emptyD2}
The last instruction of $\mathcal{D}^2 \mathcal{I}$ can be executed only if $D_2$ is empty.
\end{cor}

\emph{Proof.}\quad The previous lemma tells that condition ($\alpha$) is always true,
so instruction 5 of $\mathcal{D}^2 \mathcal{I}$ can always be executed provided that $D_2$ is not empty.\cvd

\begin{lemma}\label{beta}
At every step, we have $Top (D_1 )<Top (I)$.
\end{lemma}

\emph{Proof.}\quad The proof works by induction, exactly in the same way as Lemma \ref{alfa}.
However, it is worth giving the details in at least one case.
Suppose that, at step $n$ of the algorithm, we have $Top (D_1 )<Top (I)$ and we perform instruction 5,
that is we move $Top(D_2)$ into $I$.
Notice that, at step $n$, we must have $Top (D_1)<Top (D_2)$, otherwise condition ($\beta$) would hold,
and so instruction 3 would be performed by $\mathcal{D}^2 \mathcal{I}$ instead of instruction 5.
Therefore, at step $n+1$, we have $Top (D_1 )<Top (I)$,
because $Top(I)$ at step $n+1$ is exactly $Top(D_2 )$ at step $n$.\cvd

\begin{cor}\label{emptyD1}
The last instruction of $\mathcal{D}^2 \mathcal{I}$ can be executed only if $D_1$ is empty.
\end{cor}

\emph{Proof.}\quad We know from Corollary \ref{emptyD2} that
$D_2$ must be empty in order to execute instruction 6.
If $D_1$ were not empty, then condition ($\beta$) would be satisfied, thanks to the previous lemma,
and so instruction 3 would be performed.\cvd

From now on, we aim at showing that,
if $\pi$ is a 2-sortable permutation, then there exists a sorting algorithm for $\pi$
which has many properties that also $\mathcal{D}^2 \mathcal{I}$ has.
In the end, we will prove that such properties do characterize algorithm $\mathcal{D}^2 \mathcal{I}$.

\begin{prop}
Let $\pi$ be a 2-sortable permutation.
There exists a sorting algorithm for $\pi$ which performs operation $d_0$ (resp., $d_1 ,d_2$) only if
condition ($\gamma$) (resp., ($\beta$), ($\alpha$)) holds.
\end{prop}

\emph{Proof.}\quad Condition ($\alpha$) is obviously necessary in order to perform $d_2$, since $I$ is an increasing stack.

Consider now condition ($\beta$). Again, in order to perform $d_1$ we must have $Top(D_2 )<Top(D_1 )$, since $D_2$ is a decreasing stack.
Moreover, we will show that it is necessary to have $Top(D_1 )<Top(I)$ if we want to perform $d_1$ and eventually sort the input.
Indeed, suppose that $Top(I)<Top(D_1 )$ and set $Top(I)=b$, $Top(D_2 )=a$ and $Top(D_1 )=c$. There are two cases to analyze.
If $a<b$, then performing $d_1$ would force $b$ to reach the output before $a$, which would cause the sorting process to fail.
On the other hand, if $b<a$, we must have that $b$ is the next element to be output.
Therefore we can perform $d_3$ until $Top(I)$ is not the next element to be output.
But in this case necessarily $a<Top(I)$, and we are thus led to the previous case.

Finally, we analyze condition ($\gamma$).
The inequality $Top(D_1 )<Input$ is necessary in order to perform $d_0$, since $D_1$ is decreasing;
the inequality $Input<Top(I)$ is necessary as well, by an argument similar to that employed for condition ($\beta$).
We will now show that requiring the third constraint of ($\gamma$) to perform $d_0$ does not prevent the procedure to sort the input.
Suppose that the third constraint of ($\gamma$) is not satisfied and set $x=Top(D_2 )$.
This means that currently the input consists of a (nonempty) increasing sequence of elements smaller than $x$
whose last term (call it $b$) is bigger than the next one (call it $a$).
Of course, it is $a\leq x$ as well.
First of all, if it were possible to perform $d_1$, then necessarily $Top(D_2 )<Top(D_1 )$;
since we are supposing to be able to perform $d_0$, we already know that $Top(D_1 )<Input$, thus we would have $Top(D_2 )<Input$;
this would imply that the third constraint of ($\gamma$) is satisfied, which is not.
If we decide to perform $d_0$, we still cannot perform $d_1$ of course, so we can continue to perform $d_0$ until we reach $a$.
At that point, the only possible operation to perform would be $d_2$.
However, the same configuration could have been reached by performing $d_2$ before starting executing $d_0$.
This essentially means that
the set of configurations that are reachable by performing $d_2$ whenever the third constraint of ($\gamma$) is not satisfied
is a superset of the set of configurations that are reachable by performing $d_0$ in the same situation.
Thus, if the input is 2-sortable, then it is 2-sortable also by an algorithm which executes $d_0$ only if ($\gamma$) is satisfied.\cvd

At this point, it is convenient to make a brief recap.
What we have shown until now is that, if $\pi$ is a 2-sortable permutation,
then there exists a sorting algorithm for $\pi$ having the following features:
\begin{itemize}

\item if $Top(I)$ is the next element to be output, it performs $d_3$;

\item it executes instruction 2 of $\mathcal{D}^2 \mathcal{I}$ whenever it is possible to execute it;

\item it performs operation $d_0$ (resp., $d_1 ,d_2$) only if condition ($\gamma$) (resp., ($\beta$), ($\alpha$)) hold;

\item if no other operation is allowed, it performs $d_3$.

\end{itemize}

In order to conclude our proof, we now need to show that, if $\pi$ is 2-sortable,
then there exists a sorting algorithm $\mathcal{ALG}$ for $\pi$ which satisfies the above listed properties and,
in addition, performs operations $d_0 ,d_1 ,d_2$ in exactly the same order as algorithm $\mathcal{D}^2 \mathcal{I}$ does.
This would mean precisely that $\mathcal{ALG}$ coincides with $\mathcal{D}^2 \mathcal{I}$, as desired.

We start by comparing operations $d_1$ and $d_2$.
From now on,
any sorting algorithm having the properties listed above will be called \emph{special},
and we will denote a generic special algorithm with $\mathcal{ALG}$ .

\begin{prop}\label{d1d2}
Let $\pi$ be a 2-sortable permutation.
There exists a special sorting algorithm $\mathcal{ALG}$ for $\pi$ for which $^{(\beta)}d_1 \, \rhd \, ^{(\alpha)}d_2$.
\end{prop}

\emph{Proof.}\quad Suppose that, at a certain point of the execution of $\mathcal{ALG}$ on $\pi$, it is possible to perform both $d_1$ and $d_2$.
Clearly, we can suppose that both instruction 1 and 2 of $\mathcal{D}^2 \mathcal{I}$ cannot be executed by $\mathcal{ALG}$.
This implies that there must exist an element $a$ of $\pi$ still in the input, which is smaller than $Top(D_2 )$.
Set $x=Top(D_2 )$ and $y=Top(D_1 )$.
If we perform $d_2$, then we would have $x=Top(I)<Top(D_1 )=y$ (since we are supposing that it was possible to perform also $d_1$).
This means that $a$ could overcome $y$ only when $y$ is already inside $I$, and this can happen only if $x$ has already been output.
This however would cause the output to be unsorted, since in the output $x$ would come before $a$, and $x>a$.
We can thus conclude that, in the hypothesis of the proposition, performing $d_2$ would make the sorting process fail,
and so $^{(\beta)}d_1 \, \rhd \, ^{(\alpha)}d_2$, as desired.\cvd

We can now observe that, if $\pi$ is a 2-sortable permutation,
then there exists a special sorting algorithm $\mathcal{ALG}$ for $\pi$ such that
Lemmas \ref{alfa} and \ref{beta} and Corollaries \ref{emptyD2} and \ref{emptyD1} hold.
In fact, all the proofs of the above mentioned results do not depend on the specific algorithm $\mathcal{D}^2 \mathcal{I}$,
except for Lemma \ref{beta}, where it is explicitly used that fact that $^{(\beta)}d_1 \, \rhd \, ^{(\alpha)}d_2$.
However, in view of the previous proposition,
without loss of generality we can assume that there is a special sorting algorithm for $\pi$ which satisfies such a condition.
In what follows, a special sorting algorithm with this additional property will be called \emph{extraspecial}
(and still denoted $\mathcal{ALG}$).

\bigskip

Before concluding our \emph{tour de force}, we still need a final preparatory result.

\begin{prop}\label{231}
A permutation $\pi$ is 2-sortable if and only if
it does not contain any occurrence $bca$ of the pattern 231 such that,
at some step of any extraspecial sorting algorithm for $\pi$,
we have $b=Top(I)$ and $c$ and $a$ are still in the input.
\end{prop}

\emph{Proof.}\quad Suppose that $\pi$ is 2-sortable and that $bca$ is an occurrence of 231 in $\pi$.
Moreover, suppose that, at some point of the extraspecial sorting algorithm $\mathcal{ALG}$,
we have $b=Top(I)$ and $c$ and $a$ are still in the input.
Then, if we continue the execution of $\mathcal{ALG}$, since the first two stacks are decreasing,
$a$ can overcome $c$ only inside the increasing stack;
but $c$ can enter the increasing stack only if $b$ is in the output.
This will cause $b$ to be output before $a$, and so the input permutation would eventually not be sorted,
which is a contradiction.

On the other hand, suppose that $\pi$ is not 2-sortable and let $\mathcal{ALG}$ be any extraspecial algorithm.
Since $\pi$ is not 2-sortable, at some point $\mathcal{ALG}$ output an element $y$ which is not the correct one;
in other words, there exists $x<y$ which is still inside one of the decreasing stacks or in the input.
However, the decreasing stacks must be empty, as a consequence of Corollaries \ref{emptyD2} and \ref{emptyD1},
hence $x$ must be in the input.
Moreover, if the $z$ is the first element of the input when $y$ goes to the output, then necessarily $z>y$,
since otherwise condition ($\gamma$) would be satisfied
(which is not possible, since $\mathcal{ALG}$ executes instruction 6).
Thus, in particular, $z\neq x$,
and the elements $yzx$ constitute an occurrence of 231 in $\pi$ which violates the required condition.\cvd

We are finally ready to conclude our proof of the optimality of $\mathcal{D}^2 \mathcal{I}$.

\begin{theorem}
The sorting algorithm $\mathcal{D}^2 \mathcal{I}$ is optimal, i.e. it sorts all 2-sortable permutations.
\end{theorem}

\emph{Proof.}\quad Let $\pi$ be a sortable permutation.
Then there exists an extraspecial algorithm $\mathcal{ALG}$ which sorts $\pi$.
The only possibility for $\mathcal{ALG}$ to be different from $\mathcal{D}^2 \mathcal{I}$
is that the order in which $\mathcal{ALG}$ performs operations $d_0 ,d_1$ and $d_2$ may be different.
However, we already know that, for an extraspecial algorithm, $^{(\beta)}d_1 \, \rhd \, ^{(\alpha)}d_2$.
What remains to do is to compare $d_0$ with $d_2$ and $d_0$ with $d_1$.

First, suppose that $\mathcal{ALG}$ is in a certain configuration, in which both $d_0$ and $d_2$ can be performed.
We can further assume that condition ($\beta$) is not satisfied, otherwise $d_2 $ would certainly not be performed,
as a consequence of Proposition \ref{d1d2}.
Set $c=Top(I), b=Top(D_2 )$ and $a=Top(D_1 )$, and call $y$ the first element of the current input which is greater than $b$ (if it exists).
Since we are supposing that condition ($\gamma$) is satisfied, the sequence from the beginning of the current input to $y$ is increasing.
If there were an element $x<b$ following $y$, then performing $d_2$ would prevent to successfully sort the permutation,
as a consequence of Proposition \ref{231} (the three elements $b,y$ and $x$ would constitute the ``bad" occurrence of 231).
Therefore, also keeping in mind that $b>a$
(since $a<c$ as a consequence of condition ($\gamma$) and we are supposing that condition ($\beta$) is not satisfied),
we can assert that the set of all numbers contained in $D_1, D_2$ and in the input before $y$ (if such an element exists) is precisely
the set of all numbers $\leq b$ which are not already in the output.
It is now possible to show that, using algorithm $\mathcal{D}^2 \mathcal{I}$, such numbers reach the output before any other number makes any move.
Indeed, $\mathcal{D}^2 \mathcal{I}$ performs $d_0$ and pushes the first number of the current input inside $D_1$ (above $a$).
Then the algorithm keeps performing $d_0$ until $y$ is reached
(in fact condition ($\beta$) keeps failing to be satisfied, since all numbers before $y$ in the input are $<b$);
at this point, $D_1$ and $D_2$ contains precisely the next elements to be output, so $\mathcal{D}^2 \mathcal{I}$ performs instruction 2.
We can thus conclude that, in the considered configuration, using algorithm $\mathcal{D}^2 \mathcal{I}$ does not prevent the permutation to be sorted,
hence performing $d_0$ instead of $d_2$ is irrelevant (if not necessary).

Now suppose that $\mathcal{ALG}$ is in a certain configuration, in which both $d_0$ and $d_1$ can be performed.
Letting $Top(I)=d$, $Top(D_2 )=a$, $Top(D_1 )=b$ and $Input=c$, we then know that $a<b<c<d$.
If $\mathcal{ALG}$ chose to perform $d_0$, then $c$ would be pushed into $D_2$, with $b$ still in the same stack.
Clearly, sooner or later, there would be a step of $\mathcal{ALG}$ moving $b$ from $D_2$ to $D_1$.
Let us now focus on this exact moment (when $b$ is pushed into $D_1$) and call the resulting configuration $\aleph$:
we claim that,
if we modify $\mathcal{ALG}$ by just performing $d_1$ instead of $d_0$ in the configuration described at the beginning of the present paragraph,
we can reach the same configuration $\aleph$ mentioned above.
So suppose that, after having performed $d_0$ and before moving $b$ to $D_2$,
the elements that $\mathcal{ALG}$ has pushed into $D_1$ are $c,c_1 ,\ldots ,c_k$.
Clearly, when $b$ is moved into $D_2$, such elements must all be inside $I$, since they are all greater than $b$ and $D_2$ is decreasing.
If, in the meanwhile, $a$ has not been pushed into $I$, then we can reach the same configuration by first moving $b$ into $D_2$
(thus performing $d_1$) and then moving all the elements $c,c_1 ,\ldots ,c_k$ into $I$ by performing the same sequence of operations.
Otherwise, if $a$ would have been moved into $I$ before all elements $c,c_1 ,\ldots ,c_k$ reach $I$
(possibly together with some further elements from $D_2$),
this should have been done in a configuration in which both $d_0$ and $d_1$ were not legal
(since we have already shown that both $d_0 \rhd d_2$ and $d_1 \rhd d_2$).
This is however impossible, since we will now see that $d_1$ is certainly legal.
Indeed, focussing on the instant immediately before $a$ is pushed into $I$, since $b$ is into $D_1$,
$Top(D_1 )>b$ (since $D_2$ is decreasing) and we know that $b>a$, hence $Top(D_2 )<Top(D_1 )$.
Moreover, since $\mathcal{ALG}$ is extraspecial, we also know from Lemma \ref{beta} that $Top(D_1 )<Top(I)$.
Therefore condition $\beta$ is satisfied, hence $d_1$ is legal.
Summing up, we have shown that, if both $d_0$ and $d_1$ are legal,
then performing $d_0$ leads to a configuration which can be reached also performing $d_1$ instead.
As a consequence, performing $d_1$ instead of $d_0$ preserves sortability.\cvd

The sequence counting permutations of length $n$ that are sortable using the $\mathcal{D}^2 \mathcal{I}$ machine starts
1,1,2,6,24,117,651,3961,25661,174062,1222784, and appears to be new to \cite{Sl}.

\section{Some further algorithms}

As we have seen in the previous section,
there exists an optimal algorithm for the $\mathfrak{D}^2 \mathfrak{I}$ machine which is able to sort all sortable permutations.
However, it is not a very easy one:
in order to understand which operation should be performed at each step, one needs to check certain conditions,
which in some cases are rather weird.
Another approach could be to consider some much easier algorithms, which of course fail to be optimal,
but have the nice feature of being more intuitive.

In the present section we briefly sketch two very natural algorithms,
one of which turns out to be ``too easy" whereas the other one reveals to be ``too hard".

\subsection{A left-greedy algorithm}

Our first proposal is a left-greedy procedure for the $\mathfrak{D}^k \mathfrak{I}$ machine:
at each step, we perform the operation $d_j$ having maximum index $j$ among the legal available operations.
In other words, such a left-greedy procedure is characterized by the following chain of priorities:
$$
d_{k+1}\rhd d_k \rhd d_{k-1}\rhd \cdots d_1 \rhd d_0 .
$$

Setting $Sort^{(lg)}_k = \{ \pi: \pi \mbox{ is sorted by the left-greedy procedure} \}$,
it turns out that $Sort^{(lg)}_k$ is in fact a class which we are able to characterize completely.
The choice of a left-greedy strategy, instead of a right-greedy one, is suggested by the results contained in \cite{SM2}.

\begin{prop}
For every $k\geq 0$, $Sort^{(lg)}_k$ is a class with basis $\left\lbrace 231 \right\rbrace$.
\end{prop}

\emph{Proof.}\quad We start by proving that if $\pi$ contains $231$, then $\pi \notin Sort^{(lg)}_k$.
Let $bca$ be an occurrence of the pattern $231$ in $\pi$.
If $b$ is pushed into $I$ before $c$, then $\pi$ cannot be sorted, as a consequence of Lemma \ref{not_sort2}.
Then suppose that $b$ is stuck into a decreasing stack $D_j$, for some $j$.
In particular, since the algorithm is left-greedy, this implies that $D_k$ is not empty
(more precisely, each stack $D_i$, with $i\ge j$, has to contain at least one element).
Let $z$ be the first element that reaches $D_k$ without going directly into $I$ and consider the step in which $z$ is pushed into $D_k$;
again because we are using a left-greedy strategy, the next stack $I$ cannot be empty at that moment.
Let $y=Top(I)$. Note that $y<z$, otherwise $z$ would be pushed into $I$.
Moreover, since $y$ is not pushed into the output, there must still be an element $t<y$ that is not in the output (and neither in $I$, of course).
In particular, $t$ follows $z$, because $z$ is the top of $D_k$.
We are thus in a position to apply Lemma \ref{not_sort2} with the three elements $t<y<z$, which is enough to conclude that $\pi \notin Sort^{(lg)}_k$.

Conversely, we have to show that, if $\pi \notin Sort^{(lg)}_k$, then $\pi$ contains the pattern $231$.
Factorize $\pi$ as $\pi =\alpha_1 \alpha_2 \cdots \alpha_r$, where each $\alpha_i$ is a maximal decreasing sequence.
W.l.o.g., we can suppose that, if $\alpha_1$ contains $i$ elements, then $\alpha_1 \neq i(i-1)\cdots 21$;
otherwise, in fact, we could simply remove $\alpha_1$ and consider the remaining permutation:
since by hypothesis $\pi$ is not sortable,
there must be an index $h$ such that $\alpha_h$ is not the set of the next elements to be output.
So suppose that $\alpha_1 =\pi_1 \pi_2 \cdots \pi_i \neq i(i-1)\cdots 21$, hence $\pi_i <\pi_{i+1}$.
All the elements of $\alpha_1$ are pushed into the increasing stack, whereas $\pi_{i+1}$ remains stuck into $D_k$.
Notice that the hypothesis on $\alpha_1$ implies that not all elements inside the increasing stack can be output,
since there is at least one element $x$ following $\pi_{i+1}$ in $\pi$ which is smaller than all elements of $\alpha_1$.
Such an element $x$ is still in the input when $\pi_{i+1}$ reaches $D_k$ (since all the remaining decreasing stacks are clearly empty).
Call $y$ the top of the increasing stack when $\pi_{i+1}$ reaches $D_k$:
then the three elements $y,\pi_{i+1}$ and $x$ are an occurrence of the pattern 231 in $\pi$.
\cvd

As a consequence of the previous proposition, our left-greedy procedures sort precisely the same permutations as Stacksort does.
Thus, in a sense, adding any number of decreasing stacks before an increasing one does not improve the sorting power of the machine,
provided that we always perform the leftmost legal operation.
This does not mean, however, that the left-greedy algorithms are equivalent to Stacksort.
Indeed, taking for instance $k=1$ and the input permutation 2341, the left-greedy $\mathfrak{D}^k \mathfrak{I}$ machine returns 2134 as output, whereas Stacksort returns 2314.
In other words, while the preimage of the identity permutation is the same for Stacksort and for every left-greedy $\mathfrak{D}^k \mathfrak{I}$ machine,
the preimages of other permutations are in general different.
It would be certainly interesting to investigate more deeply the preimage of a generic permutation for the left-greedy $\mathfrak{D}^k \mathfrak{I}$ machine.

%

\subsection{A quasi left-greedy algorithm.}

There is a better way to design an algorithm which is quasi left-greedy and is able to sort more permutations than the previous one.
The idea is to give the increasing stack a privileged role, using it only when no other operation is possible.
Formally, at each step we choose to perform the first legal operation according to the following priority rule:

\begin{center}
$d_{k+1} \rhd d_{k-1} \rhd d_{k-2} \rhd \cdots \rhd d_1 \rhd d_0 \rhd d_k$.
\end{center}

This quasi left-greedy procedure is similar to the optimal algorithm for the $\mathfrak{D}^2 \mathfrak{I}$ machine described in Section \ref{optimal},
the only difference being that no additional conditions are required in order to perform operations
(other than the fact that each operation can be performed only if it is legal, of course).

In analogy with the previous case,
define $Sort^{(qlg)}_k$ to be the set of permutations sorted by the quasi left-greedy algorithm with $k$ decreasing stacks;
such permutations will be called \emph{qlg-$k$-sortable permutations}.
We observe immediately that the permutation $231$ is qlg-2-sortable. 
Unfortunately, $Sort^{(qlg)}_k$ is not in general a permutation class, except for the case $k=1$,
for which we have the following result, whose proof can be found in \cite{CCF}.

\begin{lemma}
$Sort^{(qlg)}_1$ is a class with basis $\left\lbrace 213 \right\rbrace$.
\end{lemma}

When $k>1$ things become much more involved.
As an example, for $k=2$, the permutation $631425$ is qlg-2-sortable, whereas its subpermutation $52314$ is not.
In fact, a complete characterization of $Sort^{(qlg)}_2$ appears to be quite hard.
In the rest of the section we will prove 
some partial results that should make abundantly clear that
understanding the set of qlg-$k$-sortable permutations is a very hard task.

\begin{prop}{\label{Charact}}
Let $\pi$ be a qlg-2-sortable permutation. Then
\begin{itemize}
\item $\pi$ avoids $3214$;
\item 
if $\pi$ contains the pattern $52314$, then each occurrence of $52314$ can be extended to one of the following patterns, where the additional elements are marked with a dot:
\begin{itemize}
\item $63 \dot{1}425$;
\item $7\dot{2}\dot{1}4536$, $7\dot{3}\dot{1}4526$;
\item $\dot{7}\dot{2}8\dot{1}4536$, $\dot{7}\dot{3}8\dot{1}4526$;
\item $\dot{8}\dot{2}7\dot{1}4536$, $\dot{8}\dot{3}7\dot{1}4526$.
\end{itemize}
\end{itemize}

\end{prop}

\emph{Proof.} \quad We start by proving that any sortable permutation $\pi$ cannot contain the pattern $3214$. Suppose that $cbad$ is an occurrence of $3214$ in $\pi$ and let $m$ be the smallest element that follows $b$ and precedes $d$. We focus on the instant when $d$ is pushed into $D_1$. Notice that:
\begin{itemize}
\item $c$ has to be contained in $I$, because $c>b>a$ and the stacks $D_1$ and $D_2$ are decreasing;
\item $m$ is still in $D_1$; in fact it cannot go directly into $D_2$ because there are elements in $D_2$ which are larger than it (at least $b$ or the elements that replaced it). Moreover it is the smallest element before $d$, so the next element of the input cannot force $m$ to be pushed into $D_2$.
\end{itemize}
Therefore we can apply Lemma \ref{not_sort2} with the elements $c,d$ and $m$ and conclude that $\pi$ cannot be sorted.

We now consider the pattern $52314$. Let $ebcad$ be an occurrence of $52314$ in $\pi$. Without loss of generality, we can suppose that $e$ is the rightmost element of $\pi$ preceding $b$ which plays the role of 5. 
In fact, given any other occurrence $\hat{e}bcad$ of $52314$, with $\hat{e}$ to the right of $e$, any extension of such an occurrence to one of the desired patterns would also give a similar extension of $ebcda$. In other words, we can suppose that there is no element greater than $e$ between $e$ and $b$ in $\pi$. The fact that $\pi$ is sortable, together with Lemma \ref{not_sort2}, guarantees that, when $d$ is pushed into $D_1$, one of the two following configurations holds:
\begin{enumerate}
\item $a,b$ and $c$ are all contained in the increasing stack $I$;
\item $a$ is contained in $D_1$, while $b$ and $c$ are contained in $D_2$.
\end{enumerate}
In the first case, when $a$ is pushed into $I$ (with $d$ still in the input, of course), $D_1$ has to be nonempty and the next element of the input $z$ has to be smaller than $a$. So the elements $b,a,z,d$ form an occurrence of the pattern $3214$, which is a contradiction with what we have just proved above.

We now focus on the second case. When $d$ is pushed into $D_1$, we have:

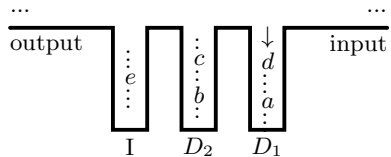
\begin{center}
\begin{threestacks}
\fillthreestacks{...}{...e...}{..b..c..}{...a...d {\downarrow}}{...}
\end{threestacks}
\end{center}

Suppose there is an element $x$ between $b$ and $c$ in $\pi$ such that $x < b$. If $x>a$, then $bxad$ is an occurrence of $3214$, which is again a contradiction. If $x<a$, we have that $ebxcad$ is an occurrence of $63\dot{1}425$, as desired. Otherwise, suppose that $x>b$ for each $x$ between $b$ and $c$ in $\pi$. This implies that $b$ is pushed directly into $D_2$ by the algorithm, because $c$ lies above $b$ in $D_2$ and no other element can push $b$ into $D_2$ before $c$ enters. As a consequence, $e$ must be already in $I$ when $b$ is pushed into $D_1$, thus, when $e$ is pushed into $I$, setting $y_1=top(D_1)$ and denoting with $y_2$ the next element of the input, we have $e > y_1 > y_2$. Moreover, it cannot be $y_2=b$, otherwise $b$ could not go directly into $D_2$, because it would be blocked by the smallest element $t$ inside $D_1$ which is greater than $y_2$ (such an element exists since $y_1>y_2$).

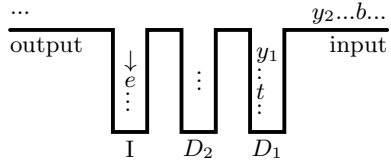
\begin{center}
\begin{threestacks}
\fillthreestacks{...}{...e {\downarrow}}{...}{..t..{y_1}}{{y_2}...b...}
\end{threestacks}
\end{center}

We are now left with two distinct cases: $e$ either precedes or follows $y_1$.
\begin{enumerate}
\item If $e$ precedes $y_1$, we have the pattern $e y_1 y_2 b c a d$. Note that $y_1 <d$ as a consequence of our choice of $e$, and also $y_1<b$, otherwise $y_1 b a d$ would be an occurrence of $3214$. Therefore we have the following possibilities:
\begin{itemize}
\item $y_1>y_2>a$, hence $y_1 y_2 ad$ is an occurrence of $3214$, against the fact that $\pi$ is sortable;
\item $y_1>a>y_2$, hence $e y_1 y_2 b c a d$ is an occurrence of $7\dot{3}\dot{1}4526$.
\item $a>y_1>y_2$, hence $e y_1 y_2 b c a d$ is an occurrence of $7\dot{2}\dot{1}4536$.
\end{itemize}
\item If $e$ follows $y_1$, a thorough case by case analysis, similar to the previous one, leads to the remaining four patterns $\dot{7}\dot{2}8\dot{1}4536$, $\dot{7}\dot{3}8\dot{1}4526$, $\dot{8}\dot{2}7\dot{1}4536$ and $\dot{8}\dot{3}7\dot{1}4526$. 
\cvd
\end{enumerate}

%

The above proposition cannot be inverted,
since there exist permutations that are not qlg-2-sortable, yet satisfy the two conditions listed above.
An example is given by $11 \ 2 \ 10 \ 1 \ 4 \ 9 \ 3 \ 6 \ 7 \ 5 \ 8$;
notice, in particular, that it contains three occurrences of 52314 and each of them can be extended to one of the above barred patterns
(more specifically, two of the occurrences can be extended to $\bar{8}\bar{2}7\bar{1}4536$,
whereas the remaining one can be extended to $7\bar{2}\bar{1}4536$).

In fact, starting from the permutation $52314$,
it is possible to construct a sequence of permutations of increasing lengths whose sortability depends on the parity of the length.
To be more precise, for $m \ge 1$, define the permutation $\gamma_m \in S_{3m+2}$ as follows:
$$\gamma_m=3m+2,\ \underbrace{2, \ 3m+1, \ 1,}_{P_1}\ \underbrace{4, \ 3m, \ 3,}_{P_2} \dots ,\underbrace{2m-2, \ 2m+3, \ 2m-3,}_{P_m-1} \ \underbrace{2m, \ 2m+1, \ 2m-1,}_{P_m} 2m+2.$$
In other words, starting from $\gamma_1=52314$,
$\gamma_{m}$ is obtained by inserting a new occurrence $P_1=2,3m+1,1$ of the pattern $231$ between the first and the second element of $\gamma_{m-1}$,
then suitably rescaling the remaining elements.
We have the following result:

\begin{prop}
\begin{enumerate}
\item $\gamma_i$ is a pattern of $\gamma_{i+1}$, for each $i \ge 1$.
\item $\gamma_i \in Sort^{(qlg)}_2$ if and only if $i$ is even.
\end{enumerate}
\end{prop}

\begin{figure}[!h]
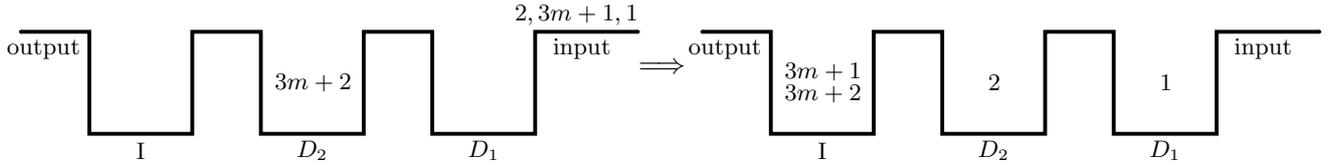

\begin{minipage}{7cm}
\begin{threestacksBig}
\fillthreestacksBig{}{}{{3m+2}}{}{2, {{3m+1}}, 1}
\end{threestacksBig}
\end{minipage}
\hspace{0.5cm}
$\Longrightarrow$
\hspace{0.5cm}
\begin{minipage}{7cm}
\begin{threestacksBig}
\fillthreestacksBig{}{{3m+2}{3m+1}}{2}{1}{}
\end{threestacksBig}
\end{minipage}

\vspace{0.5cm}
\caption{The initial stages of the quasi left-greedy algorithm with input $\gamma_m$, when $P_1$ is processed.}{\label{fig: base}}
\end{figure}

\begin{figure}
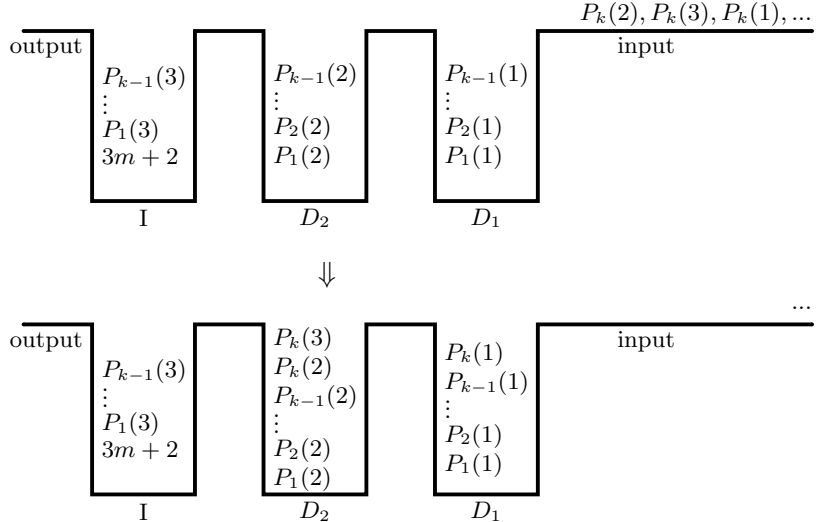

\begin{threestacksVerybig}
\hfill \fillthreestacksVerybig{}{   {3m+2}{P_1(3)}...{P_{k-1}(3)}    }{  {P_1(2)}{P_2(2)}...{P_{k-1}(2)}  }{  {P_1(1)}{P_2(1)}...{P_{k-1}(1)}  }{{P_k(2)},{P_k(3)},{P_k(1)},...}
\end{threestacksVerybig}

\vspace{1.5cm}

\hspace{6cm} $\Downarrow$

\begin{threestacksVerybig}
\fillthreestacksVerybig{}{   {3m+2}{P_1(3)}...{P_{k-1}(3)}    }{  {P_1(2)}{P_2(2)}...{P_{k-1}(2)}{P_k(2)}{P_k(3)}  }{  {P_1(1)}{P_2(1)}...{P_{k-1}(1)}{P_k(1)}  }{...}
\end{threestacksVerybig}

\vspace{1.5cm}
\caption{The behavior of the algorithm on $P_k$, when $k$ is even. Here the algorithm pushes $P_k$ into $D_1$ and $D_2$ without extracting other elements.}{\label{fig: even}}
\end{figure}

\begin{figure}
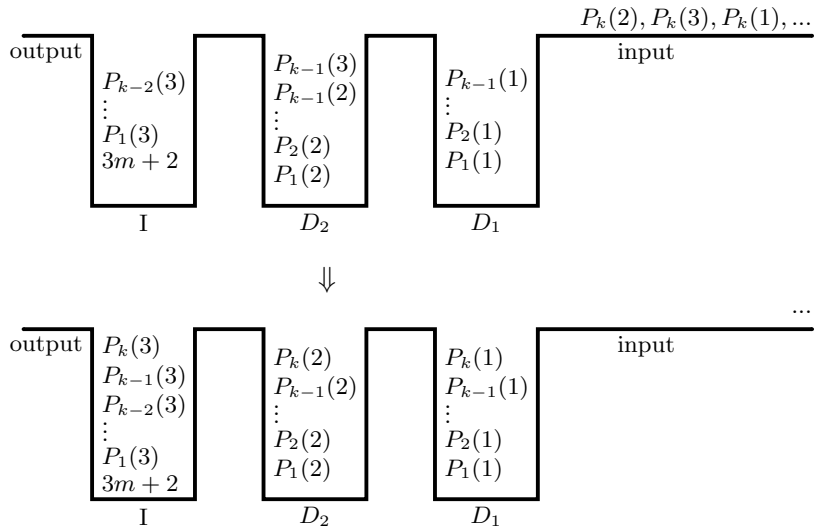

\begin{threestacksVerybig}
\hfill \fillthreestacksVerybig{}{   {3m+2}{P_1(3)}...{P_{k-2}(3)}    }{  {P_1(2)}{P_2(2)}...{P_{k-1}(2)}{P_{k-1}(3)}  }{  {P_1(1)}{P_2(1)}...{P_{k-1}(1)}  }{{P_k(2)},{P_k(3)},{P_k(1)},...}
\end{threestacksVerybig}

\vspace{1.5cm}

\hspace{6cm} $\Downarrow$

\begin{threestacksVerybig}
\fillthreestacksVerybig{}{   {3m+2}{P_1(3)}...{P_{k-2}(3)}{P_{k-1}(3)}{P_k(3)}    }{  {P_1(2)}{P_2(2)}...{P_{k-1}(2)}{P_k(2)}  }{  {P_1(1)}{P_2(1)}...{P_{k-1}(1)}{P_k(1)}  }{...}
\end{threestacksVerybig}

\vspace{1.5cm}
\caption{The behavior of the algorithm on $P_k$, when $k$ is odd. Here $P_{k-1}(3)$ and $P_k(3)$ are pushed into the increasing stack.}{\label{fig: odd}}
\end{figure}

\emph{Proof. (sketch)} \quad The first statement follows directly from the definition of $\gamma_m$. To prove the second one, we analyze how the quasi left-greedy algorithm manages the occurrences $P_k$ of the pattern $231$ in $\gamma_m$. The crucial remark is that, when $k$ is even, the elements of $P_k$ can be pushed into the decreasing stacks without extracting other elements, whereas this cannot be done when $k$ is odd. Set $P_k=P_k(2) P_k(3) P_k(1)$, for $k=1,\dots,m$. The behavior of the algorithm in both cases is represented in \figurename~\ref{fig: base}, \figurename~\ref{fig: even} and \figurename~\ref{fig: odd}.

As a consequence, it is easy to check that, if $m$ is even, then the last $4$ elements of $\gamma_m$ can be pushed into the decreasing stacks, so the permutation is eventually sorted. On the other hand, if $m$ is odd, then the second-to-last element $2m-1$ forces $2m+1$ to be pushed into the increasing stack immediately above $2m+3$, and the final element $2m+2$ will be output in the wrong position. Therefore the algorithm does not sort $\gamma_m$.
\cvd


The existence of an infinite chain of permutations which are alternately sortable and nonsortable suggests that
it should be quite difficult to obtain a simple characterization of $Sort^{(qlg)}_2$;
it is also conceivable that it should be possible to adapt the above proposition to larger values of $k$, thus obtaining similar (negative) results.

\section{Final remarks}

In the present work we started the analysis of a sorting device consisting of $k$ decreasing stacks followed by an increasing one,
generalizing the case $k=1$ addressed in \cite{SM}.
In general, the problem of characterizing sortable permutations in terms of forbidden patterns seems quite hard,
due to the fact that the basis is infinite, as shown in Theorem \ref{antichain}.
We have however been able to describe an optimal algorithm in the case $k=2$ which can sort every sortable permutation.
Such an algorithm employs a strategy which is surely nontrivial.
Thus we have also briefly discussed some simpler algorithms, which are not able to sort all sortable permutations but are certainly simpler to describe.

There are of course several items that remain to be investigated. Some of them are the following:

\begin{itemize}
\item determine the complexity of the optimal algorithm for the $\mathfrak{D}^2 \mathfrak{I}$ machine;
\item enumerate sortable permutations, both in the general case and in the restricted (left-greedy and quasi left-greedy) cases;
\item study the machine consisting of two passes through the $\mathfrak{D} \mathfrak{I}$ machine described in \cite{SM}:
are there analogies with West 2-stack-sortable permutations?
\end{itemize}


\begin{thebibliography}{Com79}

\bibitem{AL} M. Albert,\quad \emph{PermLab: Software for Permutation Patterns},\quad
at http://www.cs.otago.ac.nz/staffpriv/malbert/permlab.php.

\bibitem{BO} M. Bona,\quad \emph{A survey of stack sorting disciplines},\quad
Electron. J. Combin.,\quad 9(2) (2002-2003) \#A1.

\bibitem{Bona} M. Bona,\quad \emph{Combinatorics of Permutations},\quad
Discrete Mathematics and Its Applications, CRC Press, 2004.

\bibitem{CCF} G. Cerbai, A. Claesson, L. Ferrari,\quad \emph{Stack sorting with restricted stacks},\quad
available at https://arxiv.org/abs/1907.08142.

\bibitem{K} S. Kitaev,\quad \emph{Patterns in permutations and words},\quad
Monographs in Theoretical Computer Science. An EATCS Series. Springer, Heidelberg, 2011.

\bibitem{KN} D. E. Knuth,\quad \emph{The art of computer programming, vol. 1, Fundamental Algorithms},\quad
Addison-Wesley, Reading, Massachusetts, 1973.

\bibitem{KR} D. Kremer,\quad \emph{Permutations with forbidden subsequences and a generalized Schr\"oder number},\quad
Discrete Math.,\quad 218 (2000) 121--130.

\bibitem{Sl} N. J. A. Sloane,\quad \emph{The On-Line Encyclopedia of Integer Sequences},\quad at oeis.org.

\bibitem{SM2} R. Smith,\quad \emph{Comparing algorithms for sorting with $t$ stacks in series},\quad
Ann. Comb.,\quad 8 (2004) 113--121.

\bibitem{SM} R. Smith,\quad \emph{Two stacks in series: a decreasing stack followed by
an increasing stack},\quad Ann. Comb.,\quad 18 (2014) 359–-363.

\bibitem{TA} R. E. Tarjan,\quad \emph{Sorting using networks of queues and stacks},\quad
Journal of the ACM,\quad 19 (1972) 341--346.

\bibitem{WE} J. West,\quad \emph{Permutations with forbidden subsequences and stack-sortable permutations},\quad
PhD thesis, Massachusetts Institute of Technology, 1990.

\bibitem{WE2} J. West,\quad \emph{Sorting twice through a stack},\quad
Theoret. Comput. Sci.,\quad 117 (1993) 303--313.

\end{thebibliography}
\end{document}